\documentclass[pra,twocolumn]{revtex4}
\usepackage{graphicx}
\usepackage{color}
\usepackage{times}
\usepackage{amstext,amsmath,amssymb}
\usepackage{bbm}

\begin{document}

\title{Application of indirect Hamiltonian tomography to systems with short coherent times}

\author{Koji Maruyama$^{1},$ Daniel Burgarth$^{2}$, Akihito Ishizaki$^{3,4}$, K. Birgitta Whaley$^{3,5}$, and Takeji Takui$^{1}$}
\affiliation{$^{1}$Department of Chemistry and Materials Science, Osaka City University, Osaka, 558-8585 Japan}
\affiliation{$^{2}$Institute of Mathematics and Physics, University of Aberystwyth, Aberystwyth SY23 3BZ, UK}
\affiliation{$^{3}$Department of Chemistry, University of California, Berkeley,CA 94720, USA}
\affiliation{$^{4}$ Physical Biosciences Division, Lawrence Berkeley National Laboratory, Berkeley, CA 94720, USA}
\affiliation{$^{5}$Berkeley Quantum Information and Computation Center, University of California, Berkeley, CA 94720, USA}

\begin{abstract}
The identification of parameters in the Hamiltonian that describes complex many-body quantum
systems is generally a very hard task. Recent attention has focused on such problems of
Hamiltonian tomography for networks constructed with two-level systems. For open quantum systems,
the fact that injected signals are likely to decay before they accumulate sufficient information
for parameter estimation poses additional challenges. In this paper, we consider use of the
gateway approach to Hamiltonian tomography \cite{Burgarth2009,Burgarth2009a} to complex quantum
systems with a  limited set of state preparation and measurement probes. We classify graph
properties of networks for which the Hamiltonian may be estimated under equivalent conditions on
state preparation and measurement. We then examine the extent to which the gateway approach may be
applied to estimation of Hamiltonian parameters for network graphs with non-trivial topologies
mimicking biomolecular systems.
\end{abstract}
\maketitle

\section{Introduction}
\label{sec:intro}

Precise information about the Hamiltonian of many-body quantum systems is crucially important for
analysis and prediction of their dynamics, especially to understand the extent to which a given
subsystem behaves quantum mechanically.  If the subsystem of interest is well isolated from the
remainder, i.e., from its environment, in the sense that its dynamics is immune to the effect of
noise, then the time evolution is unitary and observable data may be considered `clean' enough to
extract good information on the subsystem Hamiltonian. However this is clearly an idealized
situation that is rarely encountered.  Furthermore, the general procedures to estimate even a
small part of the Hamiltonian that acts on a many-body system are, in general, very complex and
require a large number of measurements of different observables. This leads to a further challenge
which is that as the data acquisition process becomes more elaborate and accesses more of the
system, it is likely to introduce an increasing amount of measurement noise.

Several approaches for Hamiltonian identification, or \textit{Hamiltonian tomography}, have been
recently proposed that seek to reduce the complexity of the procedure by making use of some a
priori knowledge about the physical system
\cite{Cole2005,Cole2006,Mohseni2006,Mohseni2008,Levi2007,Schirmer2009,Ashhab2006,Oxtoby2009,Schmiegelow2011}.
One approach is to map the many-body system onto a quantum network and to make use of knowledge
about the topology of this network to devise protocols that extract desired Hamiltonian parameters
from measurements on a restricted portion of the network.  Following the demonstration that the
Hamiltonian parameters of one-dimensional chain of spin-1/2 particles may be determined by
measurements on a single spin \cite{Burgarth2009,Franco2009}, this approach has been generalized
to more general spin networks with restricted measurement access on a small \textit{gateway}
region \cite{Burgarth2009a} as well as to more general Hamiltonians \cite{Burgarth2011}.
Reference \cite{Franco2009} also showed that such an estimation scheme may be robust against noise
under weak-coupling conditions. For sparse Hamiltonians, a different approach has recently been
developed using the method of compressed sensing \cite{Shabani2011a,Shabani2011}, which has also
been applied to quantum state tomography \cite{Gross2010}. The compressed sensing approach allows
determination of both higher order Hamiltonians and system-bath interactions, but is limited to
sparse Hamiltonians. Other approaches have been developed based on convex
optimization~\cite{Young2009} and Bayesian estimation~\cite{Schirmer2009}. While these Hamiltonian
tomography approaches are related to the better known quantum process tomography (QPT)
\cite{NIELSEN,DAriano2001,Altepeter2003,Yuen-Zhou2011,Yuen-Zhou2011a} that (together with quantum
state and quantum measurement tomographies) provides a complete characterization of quantum
dynamics, they differ from QPT in seeking to reconstruct the desired parameters with a minimal
amount of resources.

In this paper we explore the use of the gateway scheme outlined in \cite{Burgarth2009a} for
determination of Hamiltonian parameters for an open quantum system under conditions of restricted
access.  The approach of Ref. \cite{Burgarth2009a} was based on the assumption of long coherence times,
which allowed the injected signal (spin wave) to go back and forth in the network many times so
that the information about spin interactions may be encoded in the signal. For dissipative
systems, we cannot in general expect such a long lifetime of the signal and it will generally be
susceptible to decay before coming back to the injection site, even though the initial time
evolution for a short time may be seen to be coherent. We therefore limit our attention here to
complex systems in which a subsystem does show such coherent short time evolution.

One prototype of this latter situation that is of considerable current interest is the subsystem
of pigments in photosynthetic light harvesting systems. Recent spectroscopic experiments have
shown that electronic energy transfer dynamics in such systems displays coherence for several
hundreds of femtoseconds
\cite{Engel2007,Panitchayangkoon2010,Collini-etal:Nature2010,Calhoun-Fleming:JPCB2009}. Although
effects of vibrational contributions are not entirely clear, these coherences are generally
accepted to reflect quantum coherences between different excitonic states that may be described by
superpositions of single molecule electronic excitations, and are thus amenable to a two-level
pseudo-spin representation. In this work we shall consider a network of pseudo-spins that mimics
pigments in a light harvesting protein.

We first review the gateway scheme of Refs. \cite{Burgarth2009,Burgarth2009a}, introducing the
graph theoretic description and notion of infection between different regions of the network
(graph) (Sec.~\ref{sec:gateway}).  We then summarize the minimal restrictions on measurement
access via spectroscopic measurements in a pigment-protein complex (Sec.~\ref{sec:measurements}).
These differ from the measurement requirements for spin networks \cite{Burgarth2009a} and thus
necessitate an extension of that approach. We present a classification of network topologies that
are accessible to Hamiltonian tomography under the current scheme. In Sec.~\ref{sec:FMO} we then
investigate the extent to which the scheme may be applied to a network graph mimicking pigments
embedded in the Fenna-Matthews-Olson protein of photosynthetic green
bacteria~\cite{Fenna-Matthews:Nature1975}. A discussion and analysis of the limitation posed by
restriction to short time scales of coherent evolution, together with indications for extensions
to remedy this, follows in Sec.~\ref{sec:discussion}.

\section{Gateway scheme of Hamiltonian tomography}
\label{sec:gateway}

In the `gateway scheme' of Hamiltonian identification \cite{Burgarth2009,Burgarth2009a}, we
consider a network of spin 1/2 pseudo-spins subject to a unitary dynamics generated by a
Hamiltonian containing pairwise interaction terms and Zeeman terms. For clarity and conciseness we
describe here only the case of excitation-conserving Hamiltonians, namely those satisfying
$[H,\sum_{n}Z_{n}]=0$, i.e., conserving the total magnetization of the pseudo-spin network.  We
further assume that all coupling strengths $c_{n}$ between spins are real and the (relative) signs
of these, but not the magnitudes, are known. We shall consider the determination of the
Hamiltonian parameters in the first excitation subspace, i.e., the subspace in which there is only
one 'up' pseudo-spin and all other pseudo-spins are `down'.  (Note that this places a restriction
on the interaction between pseudo-spins.) We illustrate the scheme here for a 1D spin chain with
nearest-neighbor interactions, for which the Hamiltonian in this subspace is given by
\begin{equation}
H_{\mathrm{1D}}=\left(\begin{array}{cccc}
b_{1} & c_{1}\\
c_{1} & b_{2} & c_{2}\\
 &  & \ddots & c_{N-1}\\
 &  & c_{N-1} & b_{N}
\end{array}\right).\label{eq:ham_1d}
\end{equation}

For a 1D chain, we start our procedure by measuring all eigenenergies $\left\{ E_{j}\right\}$ and
coefficients $\left\{ \langle E_j \vert\mathbf{1}\rangle\right\}$ for $j=1,2,\dots,N$. Here
$\lvert E_j \rangle$ are the energy eigenstates, $\lvert \mathbf{1}\rangle$ denotes
$\lvert\uparrow\downarrow\dots\downarrow\rangle=\lvert 10\dots 0\rangle$, and $N$ is the number of
pseudo-spins. Other states with a single up-spin will be denoted similarly hereafter, i.e.,
$\lvert\mathbf{n}\rangle=\lvert\downarrow\dots\uparrow\dots\downarrow\rangle$ contains only a
single spin up at the $n$-th site. We define $\rho_{n}(t)=\mathrm{tr}_{\neq
n}[U(t)\lvert\psi_0\rangle\langle\psi_0\rvert U(t)^\dagger]$ to be the reduced density matrix on
site $n$, where $U(t)=\exp(-iHt)$ is a time evolution operator for the chain.

First we initialize the state of the chain to be
$\lvert\psi_0\rangle:=1/\sqrt{2}(\lvert\mathbf{0}\rangle+\lvert\mathbf{1}\rangle)$,
namely,$1/\sqrt{2}(\lvert 0\rangle+\lvert 1\rangle)\otimes\lvert 0\dots 0\rangle$. Such an
initialization is possible by accessing only the first spin \cite{Burgarth2007c}. We then  perform
state tomography on the first spin after a time lapse $t$ to extract the reduced density matrix
$\rho_1(t)$ and repeat this at various time delays to obtain $\rho_1(t)$ as a function of time. Up
to an irrelevant phase factor, the diagonal element of $\rho_1(t)$ can be written as
\[
f_{11}(t):=\langle\mathbf{1}\vert \exp(-iHt) \vert \mathbf{1}\rangle
=
\sum_j \exp(-iE_j t) \lvert\langle E_{j}\vert\mathbf{1}\rangle\rvert^2.
\]
The eigenenergies $\{E_j\}$ and the coefficients $\{\lvert\langle E_j\vert\mathbf{1}\rangle\rvert
\}$ can then be obtained by performing the time Fourier transform of $f_{11}(t)$. Due the
arbitrariness of the global phase, we can choose all $\langle E_j|\mathbf{1}\rangle$ to be real
and positive. Detailed discussion of the factors determining the efficiency of the Fourier
transform are discussed in Ref. \cite{Burgarth2009}. As noted there, it is necessary to observe
the repeated reflections of the signal (at least $N$ times) in order to obtain an adequate signal
to noise ratio. Consequently a long time coherence is necessary for implementation of Hamiltonian
tomography with measurements only on a single spin.

With the information about $\{ E_j \}$ and $\{ \langle E_j \vert \mathbf{1}\rangle\}$ obtained
from these single spin measurements, we can proceed the parameter estimation by constructing a set
of $N^2$ equations representing $\langle E_ j \vert H_{\mathrm{1D}} \vert \mathbf{n}\rangle$ for
$1\le j,n\le N$. These equations are given by
\begin{align}
    E_j \langle E_j \vert \mathbf{1}\rangle
        & = b_1 \langle E_{j} \vert \mathbf{1}\rangle+c_1\langle E_j \vert\mathbf{2}\rangle,
        \label{eq:set_1a}
    \\
    E_j \langle E_j\vert\mathbf{n}\rangle
        & = c_{n-1} \langle E_j\vert\mathbf{n-1}\rangle+b_{n}\langle E_j\vert\mathbf{n}\rangle
    \nonumber
    \\
        &\quad + c_n\langle E_j\vert\mathbf{n+1}\rangle\quad (1<n<N),
        \label{eq:set_1b}
    \\
    E_j \langle E_j \vert \mathbf{N}\rangle
        & = c_{N-1}\langle E_j\vert\mathbf{N-1}\rangle+b_N\langle E_j\vert\mathbf{N}\rangle.
        \label{eq:set_1c}
\end{align}
Noting that $b_{1}=\langle\mathbf{1}\vert H \vert \mathbf{1}\rangle=\sum_j E_j \lvert \langle E_j
\vert \mathbf{1}\rangle\rvert^2$, each factor of which has been determined by state tomography on
spin 1, the expansion of $\lvert\mathbf{2}\rangle$ in the basis $\lvert E_j\rangle$ can be
obtained up to the constant $c_1$. The value of $c_1$ can then be obtained within a sign factor by
requiring that $\lvert\mathbf{2}\rangle$ be normalized to unity.  All other parameters $b_l,$
$c_m$ and $\langle E_j\vert\mathbf{n}\rangle$ are then subsequently obtained in the same manner
from Eq. (\ref{eq:set_1b}) (for $l=N-1, m=N$) and Eq. (\ref{eq:set_1c}). The intensities of any
local magnetic fields (which will be assumed imposed in the $z$-direction) can be then estimated
from $\left\{ b_l\right\}$.

\begin{figure}
\includegraphics[scale=0.7]{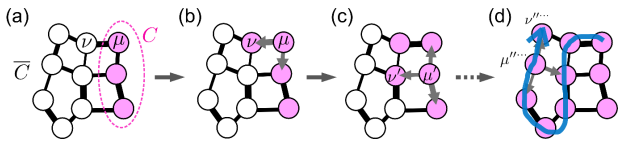}
\caption{(Color online) An example of graph infection. (a) Initially, three colored nodes in the
region $C$ are `infected'. (b) Since the node $\nu$ is the only uninfected node among the
neighbors of $\mu,$ it becomes infected as time evolves. (c) In a similar manner, $\nu^{\prime}$
becomes infected by $\mu^{\prime}$. (d) Eventually all nodes will be infected sequentially.}
\label{fig:infection}
\end{figure}

This scheme can be generalized to more complex graphs by enlarging the accessible area $C$ as was
shown in Ref. \cite{Burgarth2009a}. In that work the Hamiltonian tomography of a general graph
formed by a network of spin-1/2 systems was found to be possible if $C$ \textit{infects} the
entire graph. The infection process is defined as follows. Starting with a subset $C$ of a larger
set of nodes $V$,  suppose that all nodes in $C$ possess, i.e., are \textit{infected} with some
property. This property then spreads and infects other nodes according to the following rule: an
infected node infects a \textit{healthy} (uninfected) neighbor if and only if the latter is the
\textit{unique} healthy neighbor of the former. If eventually all nodes are infected by this
process, the initial set $C$ is referred to as an \textit{infecting subset}. Figure
\ref{fig:infection} depicts the infecting process with a simple example. We will see below that in
general, although this requirement of infection is a necessary condition of the graph, it is not
always a sufficient condition for Hamiltonian tomography under arbitrary measurements.  In
particular, we will show that there exist graphs that are not amenable to tomography under the
spectrally restricted measurement assumptions employed in the current work.

\section{Spectrally restricted Hamiltonian tomography for pseudo-spin networks}\label{sec:measurements}

We now discuss an extension of the gateway scheme for Hamiltonian tomography of a subsystem, given
access to a restricted set of spectral measurements and some short time subsystem coherence.

We consider a pseudo-spin network with XY-type interactions and local external magnetic fields, namely
\begin{eqnarray}
\label{eq:ham_XX}
H & = & \frac{1}{2}\sum_{(m,n)\in E}c_{mn}(X_{m}X_{n}+Y_{m}Y_{n})+\sum_{m\in V}b_{m}Z_{m}\nonumber \\
 & = & \sum_{(m,n)\in E}c_{mn}(\sigma_{m}^{+}\sigma_{n}^{-}+\sigma_{m}^{-}\sigma_{n}^{+})+\sum_{m\in V}b_{m}Z_{m}.
\end{eqnarray}
This defines a graph $G=(V,E)$ with $V$ the set of pseudo-spin sites and $E$ the links defined by
the spin hopping between sites.   The Hamiltonian parameters are the coupling strengths $c_{mn}$, and
the energy gaps $2b_{m}$ due to the Zeeman terms, $\sum b_{m}Z_{m}$. We employ the notation, $X_{i},Y_{i},$ and $Z_{i}$, for the
standard Pauli matrices throughout this paper: $\sigma_{m}^{\pm}=(1/2)(X_{m}+iY_{m})$ are thus the
raising and lowering operators for the $m$-th pseudo-spin.

For a network of pigments such as that considered later in this paper (Sec.~\ref{sec:FMO}), the
pseudo-spin sites are individual molecules with pseudo-spin states $|0\rangle, |1\rangle$
corresponding to the ground and first excited electronic states, and with energy gaps $2b_m$,
while the links are given by the matrix elements of coupling between transition dipole moments on
different molecules.  This corresponds to the usual Heitler-London description of excitonic
coupling between pigments \cite{Agranovich2008}. Since we restrict the analysis here to short times
during which the dynamics are coherent, we do not explicitly include other degrees of freedom here
(but see discussion in Sec.~\ref{sec:discussion}).

Given a finite window of quantum coherence of subsystem dynamics, we may develop a variant of the
gateway Hamiltonian tomography scheme via a set of spectral measurements at short times.  This is
possible with the following set of assumptions.
\begin{enumerate}
\item The network topology is known. That is, the set of interacting pairs
of sites , which plays a dominant role in the overall dynamics, is
known without precise information on the values of the coupling strengths,
$c_{mn}$.  The latter may, without loss of generality, be assumed real.
\item The sign of each $c_{mn}$ is known, but not the magnitude.

\item The energy gaps between the two pseudo-spin levels are known for the specific sites that we need to access.

\item Single site excitation is possible when we have the information on energy gaps.

\item The energy eigenvalues of the system are known.

\item Measurement in the energy eigenbasis $\left\{
|E_{j}\rangle\right\} $ is possible, i.e., the probability of finding an eigenstate
$|E_{j}\rangle$ in a state with a single site excitation $|\mathbf{n}\rangle,$ i.e., $|\langle
E_{j}|\mathbf{n}\rangle|^{2},$ can be measured.
\end{enumerate}

Assumptions 1 and 2 are the same as in the original scheme described in the previous section.
Assumption 3 is necessary for the single site excitation in Assumption 4. Assumptions 4 and 5
differ from the assumptions of the original gateway scheme of
Refs.~\cite{Burgarth2009,Burgarth2009a}, which required waiting for a signal to travel back and
forth in the chain/network (Sec.~\ref{sec:gateway}).  Since that procedure requires long coherence
times, in situations where the time over which the quantum dynamics are coherent is limited, a
\textit{global} measurement in the energy basis provides an alternative route to acquire
information about the subsystem before the excitation decays, provided that such a measurement may
be implemented on a fast enough timescale. With use of such a global measurement, the term ``to
access the site $n$'' then gains a slightly different meaning, namely ``to prepare a state'' or
``to excite the molecule'' at the site $n$, rather than to ``to measure at site $n$" as was
implicitly understood in Ref. \cite{Burgarth2009}.

The motivation for this measurement in the eigenbasis is that all necessary information for the
gateway scheme are the sets of $\{E_j\}$ and $\langle E_j|\mathbf{n}\rangle$ for all $n\in C$.
However, as we discuss below, the fact that the quantities obtained from such measurements are the
modulus of $\langle E_j|\mathbf{n}\rangle$ gives rise to modifications to the choice of sites that
should be accessed and the class of graphs to which the scheme is applicable.

\begin{figure}
\includegraphics[scale=0.75]{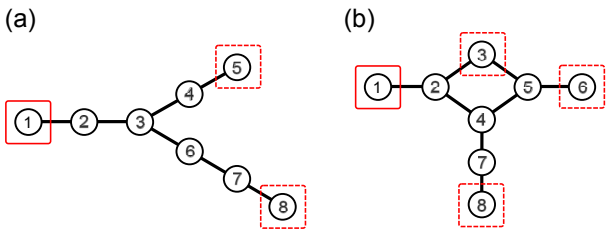}
\caption{(Color online) Two examples of graphs to which the present spectrally restricted
Hamiltonian tomography scheme can be applied. (a) For graphs with branches, the end pseudo-spin of
each branch should be accessed, i.e., this pseudo-spin should locally excited before the
measurement in the ${|E_{j}\rangle}$-basis. In this example,  sites 1, 5, and 8 (each encircled
with red lines) need to be accessed. A measurement at one of them defines the global phase, e.g.,
the site 1. The coefficients $|\langle E_{j}|\mathbf{n}\rangle|$ are measured after exciting the
red encircled sites at the other end of the graph. (b) If there is a loop in the graph, then in
addition to the end sites of branches emerging from the loop (sites 1, 6 and 8), the remaining
sites contributing to the loop (e.g., site 3, also encircled in red) need to be accessed. Note
that in both examples we need to access a larger set of sites than the smallest infecting set.
For (a), the smallest infecting set contains only two sites, e.g., 1 and 5, and for (b) three sites, such as 1, 3, and 6.}
\label{fig:branches}
\end{figure}

For 1D chains, it still suffices to access an end site, as in the original proposal in Ref.
\cite{Burgarth2009}. If the graph derived from the pseudo-spin network has branches without loops,
the end sites of all branches should be accessed.  Figure \ref{fig:branches}(a) shows an example
of such a situation. If we set the global phase by $\langle E_{j}|\mathbf{1}\rangle,$ then
measurements can only give the modulus of $\langle E_{j}|\mathbf{5}\rangle$ (and $\langle
E_{j}|\mathbf{8}\rangle)$, without their relative phases. At site 5 we have
\begin{equation}
(E_{j}-b_{5})\langle E_{j}|\mathbf{5}\rangle=c_{45}\langle E_{j}|\mathbf{4}\rangle\label{eq:branch}.
\end{equation}
Summing up the modulus squared of this equation over $j$, we can find $c_{45}^{2}.$ With the
assumed knowledge of the sign, we then obtain the value of $c_{45}$, which can then be used to
obtain the value of $|\langle E_{j}|\mathbf{4}\rangle|$. The procedure is repeated until we reach
the branching site, i.e., site 3 in Fig. \ref{fig:branches}(a), where two branches meet. The
coupling strength between sites 3 and 4 in Fig. \ref{fig:branches}(a), for instance, can be
obtained by evaluating $\sum_{j}|\cdot|^{2}$ of $(E_{j}-b_{4})\langle
E_{j}|\mathbf{4}\rangle=c_{34}\langle E_{j}|\mathbf{3}\rangle+c_{45}\langle
E_{j}|\mathbf{4}\rangle,$ resulting in $c_{34}^{2}=\sum_{j}(E_{j}-b_{4})^{2}|\langle
E_{j}|\mathbf{4}\rangle|^{2}$, from which $b_{4}$ can be obtained as before. (See text after Eq.
(\ref{eq:set_1c}).)

If there is a loop in the graph, all sites $n$ that contribute the loop need to be accessed in
order to determine $|\langle E_{j}|\mathbf{n}\rangle|.$ The necessity of knowing all $|\langle
E_{j}|\mathbf{n}\rangle|$ for the loop-forming sites $n$ derives from the requirement of having
sufficient equations to determine the couplings. If branches extrude from the loop, the access
sites should be chosen to be the end sites of these branches, just as in the case of simple graphs
having branches. (See Fig. \ref{fig:branches}(b).)

\begin{figure}
\includegraphics[scale=0.9]{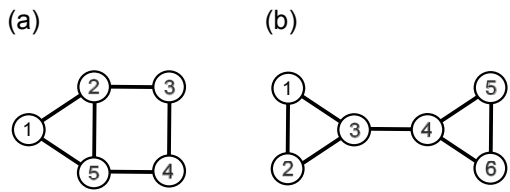}
\caption{Graphs containing more than one loop. In both examples here, the number of sites is less
than the number of edges, rendering the Hamiltonian unestimable with the spectrally restricted
approach using measurements in the energy basis.}\label{fig:loops}
\end{figure}

These examples show that the present variant of the gateway scheme for Hamiltonian tomography
cannot be applied once there are two or more loops in a connected graph. The constraint on the
available measurement given by Assumption 5 above poses a further condition on the network
topologies to which the current scheme is applicable. We illustrate this for two simple examples
of graphs in Fig. \ref{fig:loops}. Because the original gateway scheme is based on the fact that
only one unknown term, e.g., $c_{mn}\langle E_{j}|\mathbf{n}\rangle$, appears in the equation
deriving from the factor $\langle E_{j}|H|\mathbf{m}\rangle,$ we are able to obtain the value of
one new coupling strength, $c_{mn},$ by using the previously obtained knowledge on parameters for
sites other than $n$. The property of infection then guarantees that all coupling strengths can be
estimated recursively in this manner, provided that all coefficients $\langle
E_{j}|\mathbf{m}\rangle$ are known for all sites $m\in C$, as well as all eigenvalues $\{E_j\}$.
However, when there is no information on the relative phase of $\langle E_{j}|\mathbf{m}\rangle$,
the number of sites needs to be larger than or equal to the number of edges. The graphs in Fig.
\ref{fig:loops} do not fulfill this condition and as a result their Hamiltonians cannot be
estimated with the current approach.

\section{Application to a molecular network \label{sec:FMO}}

\begin{figure}
\includegraphics[scale=0.4]{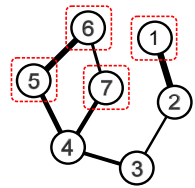}
\caption{(Color online)
An example of the set of pigment molecules to be excited in a small pigment-protein complex. Exciting those at sites 1, 5, 6 and 7 (encircled by red dashed lines)
individually is sufficient to determine all the Hamiltonian parameters. }\label{fig:fmo_graph}
\end{figure}

We now apply the gateway scheme with restricted spectral access outlined above for a network of
pseudo-spins to Hamiltonian tomography of a network graph with nontrivial topology mimicking a
pigment-protein complex. As an example, we take the geometry of the seven coupled pigments
embedded in the Fenna-Mathews-Olson protein \cite{Fenna-Matthews:Nature1975} and use the dominant
electronic couplings between pigments as in Refs. \cite{Ishizaki2009,Hoyer-etal:NJP2010}. This
leads to the network graph shown in Fig. \ref{fig:fmo_graph}.

Recent investigations have demonstrated that electronic quantum coherence in such a
pigment-protein complex persists for several hundreds of femtoseconds even at physiological
temperatures
\cite{Engel2007,Panitchayangkoon2010,Collini-etal:Nature2010,Calhoun-Fleming:JPCB2009}. This means
that application of the present Hamiltonian tomography scheme is restricted to measurements on the
timescale of a few hundred femtoseconds. We note that the presence of quantum coherence does not
necessarily imply purely unitary dynamics. If the evolution is indeed unitary, the values of
$|\langle E_j|\mathbf{n}\rangle |$ are constant in time. However, in the presence of dissipation,
the measured values of $|\langle E_j|\mathbf{n}\rangle |$ may vary. Provided that the measurement
can be made within the timescale in which the dynamics of $|E_j\rangle$ may be characterized by a
phenomenological factor $\Gamma_j$, then this time dependence would be reflected in measurement of
time dependent coefficients $|\langle E_j|\mathbf{n}\rangle |\exp(-\Gamma_j t/2)$. Measuring these
quantities at various times within the relevant timescale would then allow estimations of the
values $|\langle E_j|\mathbf{n}\rangle |$ by extrapolation back to $t=0$. Current technology
allows controlled shaping of pulses with time duration 10--20 femtoseconds, suggesting that such
an estimation might be feasible.

Using the data of $\{ E_j \} $ and $\{ \langle E_{j} \vert \mathbf{m} \rangle\,(m\in C) \}$ and
following the procedure described above with Eqs. (\ref{eq:set_1a})-(\ref{eq:set_1c}), we can then
construct the matrix elements of the symmetric matrix  corresponding to the Hamiltonian in the
one-excitation subspace.

Since the network in Fig. \ref{fig:fmo_graph} contains a loop formed by four sites, we need to
access at least four sites, 1, 5, 6, and 7, as described in Sec \ref{sec:measurements}. Let us
follow the estimation procedure again briefly for clarity. Suppose that we start from
$|\mathbf{1}\rangle$, that is, we set the global phase of $\lvert E_j\rangle$ so that $\langle E_j
\vert \mathbf{1} \rangle$ are real and positive for all $j$. Then, using Eq. (\ref{eq:set_1a}), we
have
\begin{equation}
    \sum_j \lvert (E_j-b_1) \langle E_j \vert \mathbf{1} \rangle \rvert^2
    =
    c_{12} \langle E_j \vert \mathbf{2} \rangle,
    \label{eq:1Ej}
\end{equation}
where $b_1$ can be known from $E_j$ and $\langle E_j\vert \mathbf{1}\rangle$, thus the left-hand
side of Eq. (\ref{eq:1Ej}) is equal to $\sum_j E_j^2 \langle E_j \vert \mathbf{1}
\rangle^2-b_1^2$. The estimation process then proceeds to site 4 according to Eq.
(\ref{eq:set_1b}), obtaining the values of $c_{34}$ and $\langle E_j \vert \mathbf{4} \rangle$
with a correct phase. With the measured values of $\lvert\langle\mathbf{5}\vert E_j\rangle\rvert$,
$\lvert\langle\mathbf{6}\vert E_j\rangle\rvert$, and $\lvert\langle\mathbf{7}\vert
E_j\rangle\rvert$, we can then make use of the following set of equations to obtain the coupling
strengths:
\begin{align}
     (E_j-b_4) \langle E_j \vert \mathbf{4} \rangle - c_{34} \langle E_j \vert \mathbf{3} \rangle
    & = c_{45} \langle E_j \vert \mathbf{5} \rangle + c_{47} \langle E_j \vert \mathbf{7} \rangle,
    \notag\\
     (E_j-b_5) \langle E_j \vert \mathbf{5} \rangle
    & = c_{45} \langle E_j \vert \mathbf{4} \rangle + c_{56} \langle E_j \vert \mathbf{6} \rangle,
    \notag\\
     (E_j-b_6) \langle E_j \vert \mathbf{6} \rangle
    & = c_{56} \langle E_j \vert \mathbf{5} \rangle + c_{67} \langle E_j \vert \mathbf{7} \rangle,
    \notag\\
     (E_j-b_7) \langle E_j \vert \mathbf{7} \rangle
    & = c_{67} \langle E_j \vert \mathbf{6} \rangle + c_{47} \langle E_j \vert \mathbf{4} \rangle.
    \label{eq:4eq}
\end{align}
Summing up the modulus squared of each equation over $j$ gives four equations with four unknown
parameters, $c_{45}^2$, $c_{56}^2$, $c_{67}^2$, and $c_{47}^2$. Together with the a priori
knowledge on the signs of the set $\{ c_n \}$ that was assumed initially, all coupling strengths
$c_{ij}$ can now be estimated. The remaining parameters, i.e., $b_{5}$, $b_{6}$, and $b_{7}$, may
be evaluated from $\lvert \langle E_{j}\vert\mathbf{n}\rangle\rvert$ $(n=5,6,7)$. For example,
$b_5 = \langle\mathbf{5}\rvert H \lvert \mathbf{5}\rangle = \sum_j E_j \lvert\langle E_j \vert
\mathbf{5} \rangle\rvert^2$. Thus all parameters of the Hamiltonian have now been identified,
despite the lack of the precise information about the phase of $\langle E_j \vert
\mathbf{n}\rangle$.

\section{Discussion}
\label{sec:discussion}

We have developed an extension of the gateway scheme of Hamiltonian tomography to estimation of
Hamiltonian parameters for subsystems of complex quantum systems that show coherent dynamics for a
limited period of time. We circumvent the problem of decay of quantum coherence preventing the
observation of reflections of the injected signal that was required in the original scheme of Ref.
\cite{Burgarth2009} by employing instead a measurement in the $\left\{|E_{j}\rangle\right\}
$-basis. Assuming the feasibility of such a spectrally restricted measurement, we then showed that
by choosing the right set of accessible (i.e., spectroscopically excitable) sites, the Hamiltonian
of a given network of pseudo-spins can be estimated. These constraints on measurable quantities,
in particular the lack of feasibility of measurements in the site basis, were found to modify the
requirements for the graph properties. While the Hamiltonians of one-dimensional chains are still
estimable by accessing, i.e., preparing a state at the end site only, as in the original gateway
scheme, we now find that, in general, the set of accessible sites needs to be larger than an
infecting set. Furthermore,  there exist networks to which the current estimation scheme cannot be
applied, regardless of the choice of accessible sites, because of the detailed structure of the
network topology.  This major difference results from the constraint on feasible measurements
imposed here and raises interesting questions for the interplay between network topology and
measurement capabilities in Hamiltonian identification schemes in general.

Application of this spectrally restricted Hamiltonian tomography approach to a small scale network
of molecular pigments indicated that provided the spectral measurements can be made on a timescale
significantly shorter than the characteristic time for loss of coherence, the Hamiltonian
tomography approach may yield useful estimates for parameters of the electronic Hamiltonian
describing excitonic energy transfer through a well-characterized aggregate of pigments or
pigment-protein complexes. We note nevertheless that realistic application to such systems will
require extension of the current approach to include dissipation and decoherence in a more
quantitative manner, e.g., as in Refs.
\cite{Cole2006,Schirmer2008,Franco2009,Schirmer2009,Brivio2010,Schirmer2010,Shabani2011}, to
account for the lack of unitarity that is associated with the subsystem dynamics despite the
appearance of quantum coherences at short times.

\begin{acknowledgments}
KM is grateful to the support by the JSPS Kakenhi (C) no. 22540405. KM and TT are supported in
part by Quantum Cybernetics (Grant No. 2112004), CREST-JST, and FIRST-JSPS (Quantum Information
Process). The efforts of KBW and AI are supported in part by the DARPA QuEST program.
\end{acknowledgments}

\bibliographystyle{apsrev}

\begin{thebibliography}{34}
\expandafter\ifx\csname natexlab\endcsname\relax\def\natexlab#1{#1}\fi
\expandafter\ifx\csname bibnamefont\endcsname\relax
  \def\bibnamefont#1{#1}\fi
\expandafter\ifx\csname bibfnamefont\endcsname\relax
  \def\bibfnamefont#1{#1}\fi
\expandafter\ifx\csname citenamefont\endcsname\relax
  \def\citenamefont#1{#1}\fi
\expandafter\ifx\csname url\endcsname\relax
  \def\url#1{\texttt{#1}}\fi
\expandafter\ifx\csname urlprefix\endcsname\relax\def\urlprefix{URL }\fi
\providecommand{\bibinfo}[2]{#2}
\providecommand{\eprint}[2][]{\url{#2}}

\bibitem[{\citenamefont{Burgarth et~al.}(2009)\citenamefont{Burgarth, Maruyama,
  and Nori}}]{Burgarth2009}
\bibinfo{author}{\bibfnamefont{D.}~\bibnamefont{Burgarth}},
  \bibinfo{author}{\bibfnamefont{K.}~\bibnamefont{Maruyama}}, \bibnamefont{and}
  \bibinfo{author}{\bibfnamefont{F.}~\bibnamefont{Nori}},
  \bibinfo{journal}{Phys. Rev. A} \textbf{\bibinfo{volume}{79}},
  \bibinfo{pages}{020305(R)} (\bibinfo{year}{2009}).

\bibitem[{\citenamefont{Burgarth and Maruyama}(2009)}]{Burgarth2009a}
\bibinfo{author}{\bibfnamefont{D.}~\bibnamefont{Burgarth}} \bibnamefont{and}
  \bibinfo{author}{\bibfnamefont{K.}~\bibnamefont{Maruyama}},
  \bibinfo{journal}{New J. Phys.} \textbf{\bibinfo{volume}{11}},
  \bibinfo{pages}{103019} (\bibinfo{year}{2009}).

\bibitem[{\citenamefont{Cole et~al.}(2005)\citenamefont{Cole, Schirmer,
  Greentree, Wellard, Oi, and Hollenberg}}]{Cole2005}
\bibinfo{author}{\bibfnamefont{J.~H.} \bibnamefont{Cole}},
  \bibinfo{author}{\bibfnamefont{S.~G.} \bibnamefont{Schirmer}},
  \bibinfo{author}{\bibfnamefont{A.~D.} \bibnamefont{Greentree}},
  \bibinfo{author}{\bibfnamefont{C.~J.} \bibnamefont{Wellard}},
  \bibinfo{author}{\bibfnamefont{D.~K.~L.} \bibnamefont{Oi}}, \bibnamefont{and}
  \bibinfo{author}{\bibfnamefont{L.~C.~L.} \bibnamefont{Hollenberg}},
  \bibinfo{journal}{Phys. Rev. A} \textbf{\bibinfo{volume}{71}},
  \bibinfo{pages}{062312} (\bibinfo{year}{2005}).

\bibitem[{\citenamefont{Cole et~al.}(2006)\citenamefont{Cole, Greentree, Oi,
  Schirmer, Wellard, and Hollenberg}}]{Cole2006}
\bibinfo{author}{\bibfnamefont{J.~H.} \bibnamefont{Cole}},
  \bibinfo{author}{\bibfnamefont{A.~D.} \bibnamefont{Greentree}},
  \bibinfo{author}{\bibfnamefont{D.~K.~L.} \bibnamefont{Oi}},
  \bibinfo{author}{\bibfnamefont{S.~G.} \bibnamefont{Schirmer}},
  \bibinfo{author}{\bibfnamefont{C.~J.} \bibnamefont{Wellard}},
  \bibnamefont{and} \bibinfo{author}{\bibfnamefont{L.~C.~L.}
  \bibnamefont{Hollenberg}}, \bibinfo{journal}{Phys. Rev. A}
  \textbf{\bibinfo{volume}{73}}, \bibinfo{pages}{062333}
  (\bibinfo{year}{2006}).

\bibitem[{\citenamefont{Mohseni and Lidar}(2006)}]{Mohseni2006}
\bibinfo{author}{\bibfnamefont{M.}~\bibnamefont{Mohseni}} \bibnamefont{and}
  \bibinfo{author}{\bibfnamefont{D.~A.} \bibnamefont{Lidar}},
  \bibinfo{journal}{Phys. Rev. Lett.} \textbf{\bibinfo{volume}{97}},
  \bibinfo{pages}{170501} (\bibinfo{year}{2006}).

\bibitem[{\citenamefont{Mohseni et~al.}(2008)\citenamefont{Mohseni, Rezakhani,
  and Lidar}}]{Mohseni2008}
\bibinfo{author}{\bibfnamefont{M.}~\bibnamefont{Mohseni}},
  \bibinfo{author}{\bibfnamefont{A.~T.} \bibnamefont{Rezakhani}},
  \bibnamefont{and} \bibinfo{author}{\bibfnamefont{D.~A.} \bibnamefont{Lidar}},
  \bibinfo{journal}{Phys. Rev. A} \textbf{\bibinfo{volume}{77}},
  \bibinfo{pages}{032322} (\bibinfo{year}{2008}).

\bibitem[{\citenamefont{L\'evi et~al.}(2007)\citenamefont{L\'evi, L\'opez,
  Emerson, and Cory}}]{Levi2007}
\bibinfo{author}{\bibfnamefont{B.}~\bibnamefont{L\'evi}},
  \bibinfo{author}{\bibfnamefont{C.~C.} \bibnamefont{L\'opez}},
  \bibinfo{author}{\bibfnamefont{J.}~\bibnamefont{Emerson}}, \bibnamefont{and}
  \bibinfo{author}{\bibfnamefont{D.~G.} \bibnamefont{Cory}},
  \bibinfo{journal}{Phys. Rev. A} \textbf{\bibinfo{volume}{75}},
  \bibinfo{pages}{022314} (\bibinfo{year}{2007}).

\bibitem[{\citenamefont{Schirmer and Oi}(2009)}]{Schirmer2009}
\bibinfo{author}{\bibfnamefont{S.~G.} \bibnamefont{Schirmer}} \bibnamefont{and}
  \bibinfo{author}{\bibfnamefont{D.~K.~L.} \bibnamefont{Oi}},
  \bibinfo{journal}{Phys. Rev. A} \textbf{\bibinfo{volume}{80}},
  \bibinfo{pages}{022333} (\bibinfo{year}{2009}).

\bibitem[{\citenamefont{Ashhab et~al.}(2006)\citenamefont{Ashhab, Johansson,
  and Nori}}]{Ashhab2006}
\bibinfo{author}{\bibfnamefont{S.}~\bibnamefont{Ashhab}},
  \bibinfo{author}{\bibfnamefont{J.~R.} \bibnamefont{Johansson}},
  \bibnamefont{and} \bibinfo{author}{\bibfnamefont{F.}~\bibnamefont{Nori}},
  \bibinfo{journal}{New. J. Phys.} \textbf{\bibinfo{volume}{8}},
  \bibinfo{pages}{103} (\bibinfo{year}{2006}).

\bibitem[{\citenamefont{Oxtoby et~al.}(2009)\citenamefont{Oxtoby, Rivas,
  Huelga, and Fazio}}]{Oxtoby2009}
\bibinfo{author}{\bibfnamefont{N.~P.} \bibnamefont{Oxtoby}},
  \bibinfo{author}{\bibfnamefont{A.}~\bibnamefont{Rivas}},
  \bibinfo{author}{\bibfnamefont{S.~F.} \bibnamefont{Huelga}},
  \bibnamefont{and} \bibinfo{author}{\bibfnamefont{R.}~\bibnamefont{Fazio}},
  \bibinfo{journal}{New. J. Phys.} \textbf{\bibinfo{volume}{11}},
  \bibinfo{pages}{063028} (\bibinfo{year}{2009}).

\bibitem[{\citenamefont{Schmiegelow et~al.}(2011)\citenamefont{Schmiegelow,
  Bendersky, Larotonda, and Paz}}]{Schmiegelow2011}
\bibinfo{author}{\bibfnamefont{C.~T.} \bibnamefont{Schmiegelow}},
  \bibinfo{author}{\bibfnamefont{A.}~\bibnamefont{Bendersky}},
  \bibinfo{author}{\bibfnamefont{M.~A.} \bibnamefont{Larotonda}},
  \bibnamefont{and} \bibinfo{author}{\bibfnamefont{J.~P.} \bibnamefont{Paz}},
  \bibinfo{journal}{Phys. Rev. Lett.} \textbf{\bibinfo{volume}{107}},
  \bibinfo{pages}{100502} (\bibinfo{year}{2011}).

\bibitem[{\citenamefont{Franco et~al.}(2009)\citenamefont{Franco, Paternostro,
  and Kim}}]{Franco2009}
\bibinfo{author}{\bibfnamefont{C.~D.} \bibnamefont{Franco}},
  \bibinfo{author}{\bibfnamefont{M.}~\bibnamefont{Paternostro}},
  \bibnamefont{and} \bibinfo{author}{\bibfnamefont{M.~S.} \bibnamefont{Kim}},
  \bibinfo{journal}{Phys. Rev. Lett.} \textbf{\bibinfo{volume}{102}},
  \bibinfo{pages}{054304} (\bibinfo{year}{2009}).

\bibitem[{\citenamefont{Burgarth et~al.}(2011)\citenamefont{Burgarth, Maruyama,
  and Nori}}]{Burgarth2011}
\bibinfo{author}{\bibfnamefont{D.}~\bibnamefont{Burgarth}},
  \bibinfo{author}{\bibfnamefont{K.}~\bibnamefont{Maruyama}}, \bibnamefont{and}
  \bibinfo{author}{\bibfnamefont{F.}~\bibnamefont{Nori}},
  \bibinfo{journal}{New. J. Phys.} \textbf{\bibinfo{volume}{13}},
  \bibinfo{pages}{013019} (\bibinfo{year}{2011}).

\bibitem[{\citenamefont{Shabani
  et~al.}(2011{\natexlab{a}})\citenamefont{Shabani, Kosut, Mohseni, Rabitz,
  Broome, Almeida, Fedrizzi, and White}}]{Shabani2011a}
\bibinfo{author}{\bibfnamefont{A.}~\bibnamefont{Shabani}},
  \bibinfo{author}{\bibfnamefont{R.~L.} \bibnamefont{Kosut}},
  \bibinfo{author}{\bibfnamefont{M.}~\bibnamefont{Mohseni}},
  \bibinfo{author}{\bibfnamefont{H.~A.} \bibnamefont{Rabitz}},
  \bibinfo{author}{\bibfnamefont{M.~A.} \bibnamefont{Broome}},
  \bibinfo{author}{\bibfnamefont{M.~P.} \bibnamefont{Almeida}},
  \bibinfo{author}{\bibfnamefont{A.}~\bibnamefont{Fedrizzi}}, \bibnamefont{and}
  \bibinfo{author}{\bibfnamefont{A.~G.} \bibnamefont{White}},
  \bibinfo{journal}{Phys. Rev. Lett.} \textbf{\bibinfo{volume}{106}},
  \bibinfo{pages}{100401} (\bibinfo{year}{2011}{\natexlab{a}}).

\bibitem[{\citenamefont{Shabani
  et~al.}(2011{\natexlab{b}})\citenamefont{Shabani, Mohseni, Lloyd, Kosut, and
  Rabitz}}]{Shabani2011}
\bibinfo{author}{\bibfnamefont{A.}~\bibnamefont{Shabani}},
  \bibinfo{author}{\bibfnamefont{M.}~\bibnamefont{Mohseni}},
  \bibinfo{author}{\bibfnamefont{S.}~\bibnamefont{Lloyd}},
  \bibinfo{author}{\bibfnamefont{R.~L.} \bibnamefont{Kosut}}, \bibnamefont{and}
  \bibinfo{author}{\bibfnamefont{H.~A.} \bibnamefont{Rabitz}},
  \bibinfo{journal}{Phys. Rev. A} \textbf{\bibinfo{volume}{84}},
  \bibinfo{pages}{012107} (\bibinfo{year}{2011}{\natexlab{b}}).

\bibitem[{\citenamefont{Gross et~al.}(2010)\citenamefont{Gross, Liu, Flammia,
  Becker, and Eisert}}]{Gross2010}
\bibinfo{author}{\bibfnamefont{D.}~\bibnamefont{Gross}},
  \bibinfo{author}{\bibfnamefont{Y.-K.} \bibnamefont{Liu}},
  \bibinfo{author}{\bibfnamefont{S.~T.} \bibnamefont{Flammia}},
  \bibinfo{author}{\bibfnamefont{S.}~\bibnamefont{Becker}}, \bibnamefont{and}
  \bibinfo{author}{\bibfnamefont{J.}~\bibnamefont{Eisert}},
  \bibinfo{journal}{Phys. Rev. Lett.} \textbf{\bibinfo{volume}{105}},
  \bibinfo{pages}{150401} (\bibinfo{year}{2010}).

\bibitem[{\citenamefont{Young et~al.}(2009)\citenamefont{Young, Sarovar, Kosut,
  and Whaley}}]{Young2009}
\bibinfo{author}{\bibfnamefont{K.~C.} \bibnamefont{Young}},
  \bibinfo{author}{\bibfnamefont{M.}~\bibnamefont{Sarovar}},
  \bibinfo{author}{\bibfnamefont{R.}~\bibnamefont{Kosut}}, \bibnamefont{and}
  \bibinfo{author}{\bibfnamefont{K.~B.} \bibnamefont{Whaley}},
  \bibinfo{journal}{Phys. Rev. A} \textbf{\bibinfo{volume}{79}},
  \bibinfo{pages}{062301} (\bibinfo{year}{2009}).

\bibitem[{\citenamefont{Nielsen and Chuang}(2000)}]{NIELSEN}
\bibinfo{author}{\bibfnamefont{M.~A.} \bibnamefont{Nielsen}} \bibnamefont{and}
  \bibinfo{author}{\bibfnamefont{I.~L.} \bibnamefont{Chuang}},
  \emph{\bibinfo{title}{Quantum Computation and Quantum Information}}
  (\bibinfo{publisher}{Cambridge University Press, Cambridge},
  \bibinfo{year}{2000}).

\bibitem[{\citenamefont{D'Ariano and Lo~Presti}(2001)}]{DAriano2001}
\bibinfo{author}{\bibfnamefont{G.~M.} \bibnamefont{D'Ariano}} \bibnamefont{and}
  \bibinfo{author}{\bibfnamefont{P.}~\bibnamefont{Lo~Presti}},
  \bibinfo{journal}{Phys. Rev. Lett.} \textbf{\bibinfo{volume}{86}},
  \bibinfo{pages}{4195} (\bibinfo{year}{2001}).

\bibitem[{\citenamefont{Altepeter et~al.}(2003)\citenamefont{Altepeter,
  Branning, Jeffrey, Wei, Kwiat, Thew, O'Brien, Nielsen, and
  White}}]{Altepeter2003}
\bibinfo{author}{\bibfnamefont{J.~B.} \bibnamefont{Altepeter}},
  \bibinfo{author}{\bibfnamefont{D.}~\bibnamefont{Branning}},
  \bibinfo{author}{\bibfnamefont{E.}~\bibnamefont{Jeffrey}},
  \bibinfo{author}{\bibfnamefont{T.~C.} \bibnamefont{Wei}},
  \bibinfo{author}{\bibfnamefont{P.~G.} \bibnamefont{Kwiat}},
  \bibinfo{author}{\bibfnamefont{R.~T.} \bibnamefont{Thew}},
  \bibinfo{author}{\bibfnamefont{J.~L.} \bibnamefont{O'Brien}},
  \bibinfo{author}{\bibfnamefont{M.~A.} \bibnamefont{Nielsen}},
  \bibnamefont{and} \bibinfo{author}{\bibfnamefont{A.~G.} \bibnamefont{White}},
  \bibinfo{journal}{Phys. Rev. Lett.} \textbf{\bibinfo{volume}{90}},
  \bibinfo{pages}{193601} (\bibinfo{year}{2003}).

\bibitem[{\citenamefont{Yuen-Zhou and Aspuru-Guzik}(2011)}]{Yuen-Zhou2011}
\bibinfo{author}{\bibfnamefont{J.}~\bibnamefont{Yuen-Zhou}} \bibnamefont{and}
  \bibinfo{author}{\bibfnamefont{A.}~\bibnamefont{Aspuru-Guzik}},
  \bibinfo{journal}{J. Chem. Phys.} \textbf{\bibinfo{volume}{134}},
  \bibinfo{pages}{134505} (\bibinfo{year}{2011}).

\bibitem[{\citenamefont{Yuen-Zhou et~al.}(2011)\citenamefont{Yuen-Zhou, Krich,
  Mohseni, and Aspuru-Guzik}}]{Yuen-Zhou2011a}
\bibinfo{author}{\bibfnamefont{J.}~\bibnamefont{Yuen-Zhou}},
  \bibinfo{author}{\bibfnamefont{J.~J.} \bibnamefont{Krich}},
  \bibinfo{author}{\bibfnamefont{M.}~\bibnamefont{Mohseni}}, \bibnamefont{and}
  \bibinfo{author}{\bibfnamefont{A.}~\bibnamefont{Aspuru-Guzik}},
  \bibinfo{journal}{Proc. Natl. Acad. Sci. USA} \textbf{\bibinfo{volume}{108}},
  \bibinfo{pages}{17615} (\bibinfo{year}{2011}).

\bibitem[{\citenamefont{Engel et~al.}(2007)\citenamefont{Engel, Calhoun, Read,
  Ahn, Man\v{c}al, Cheng, Blankenship, and Fleming}}]{Engel2007}
\bibinfo{author}{\bibfnamefont{G.~S.} \bibnamefont{Engel}},
  \bibinfo{author}{\bibfnamefont{T.~R.} \bibnamefont{Calhoun}},
  \bibinfo{author}{\bibfnamefont{E.~L.} \bibnamefont{Read}},
  \bibinfo{author}{\bibfnamefont{T.-K.} \bibnamefont{Ahn}},
  \bibinfo{author}{\bibfnamefont{T.}~\bibnamefont{Man\v{c}al}},
  \bibinfo{author}{\bibfnamefont{Y.-C.} \bibnamefont{Cheng}},
  \bibinfo{author}{\bibfnamefont{R.~E.} \bibnamefont{Blankenship}},
  \bibnamefont{and} \bibinfo{author}{\bibfnamefont{G.~R.}
  \bibnamefont{Fleming}}, \bibinfo{journal}{Nature}
  \textbf{\bibinfo{volume}{446}}, \bibinfo{pages}{782} (\bibinfo{year}{2007}).

\bibitem[{\citenamefont{Panitchayangkoon
  et~al.}(2010)\citenamefont{Panitchayangkoon, Hayes, Fransted, Caram, Harel,
  Wen, Blankenship, and Engel}}]{Panitchayangkoon2010}
\bibinfo{author}{\bibfnamefont{G.}~\bibnamefont{Panitchayangkoon}},
  \bibinfo{author}{\bibfnamefont{D.}~\bibnamefont{Hayes}},
  \bibinfo{author}{\bibfnamefont{K.~A.} \bibnamefont{Fransted}},
  \bibinfo{author}{\bibfnamefont{J.~R.} \bibnamefont{Caram}},
  \bibinfo{author}{\bibfnamefont{E.}~\bibnamefont{Harel}},
  \bibinfo{author}{\bibfnamefont{J.}~\bibnamefont{Wen}},
  \bibinfo{author}{\bibfnamefont{R.~E.} \bibnamefont{Blankenship}},
  \bibnamefont{and} \bibinfo{author}{\bibfnamefont{G.~S.} \bibnamefont{Engel}},
  \bibinfo{journal}{Proc. Natl. Acad. Sci. USA} \textbf{\bibinfo{volume}{107}},
  \bibinfo{pages}{12766} (\bibinfo{year}{2010}).

\bibitem[{\citenamefont{Collini et~al.}(2010)\citenamefont{Collini, Wong, Wilk,
  Curmi, Brumer, and Scholes}}]{Collini-etal:Nature2010}
\bibinfo{author}{\bibfnamefont{E.}~\bibnamefont{Collini}},
  \bibinfo{author}{\bibfnamefont{C.~Y.} \bibnamefont{Wong}},
  \bibinfo{author}{\bibfnamefont{K.~E.} \bibnamefont{Wilk}},
  \bibinfo{author}{\bibfnamefont{P.~M.~G.} \bibnamefont{Curmi}},
  \bibinfo{author}{\bibfnamefont{P.}~\bibnamefont{Brumer}}, \bibnamefont{and}
  \bibinfo{author}{\bibfnamefont{G.~D.} \bibnamefont{Scholes}},
  \bibinfo{journal}{Nature} \textbf{\bibinfo{volume}{463}},
  \bibinfo{pages}{644} (\bibinfo{year}{2010}).

\bibitem[{\citenamefont{Calhoun et~al.}(2009)\citenamefont{Calhoun, Ginsberg,
  Schlau-Cohen, Cheng, Ballottari, Bassi, and
  Fleming}}]{Calhoun-Fleming:JPCB2009}
\bibinfo{author}{\bibfnamefont{T.~R.} \bibnamefont{Calhoun}},
  \bibinfo{author}{\bibfnamefont{N.~S.} \bibnamefont{Ginsberg}},
  \bibinfo{author}{\bibfnamefont{G.~S.} \bibnamefont{Schlau-Cohen}},
  \bibinfo{author}{\bibfnamefont{Y.-C.} \bibnamefont{Cheng}},
  \bibinfo{author}{\bibfnamefont{M.}~\bibnamefont{Ballottari}},
  \bibinfo{author}{\bibfnamefont{R.}~\bibnamefont{Bassi}}, \bibnamefont{and}
  \bibinfo{author}{\bibfnamefont{G.~R.} \bibnamefont{Fleming}},
  \bibinfo{journal}{J. Phys. Chem. B} \textbf{\bibinfo{volume}{113}},
  \bibinfo{pages}{16291} (\bibinfo{year}{2009}).

\bibitem[{\citenamefont{Fenna and Matthews}(1975)}]{Fenna-Matthews:Nature1975}
\bibinfo{author}{\bibfnamefont{R.~E.} \bibnamefont{Fenna}} \bibnamefont{and}
  \bibinfo{author}{\bibfnamefont{B.~W.} \bibnamefont{Matthews}},
  \bibinfo{journal}{Nature} \textbf{\bibinfo{volume}{258}},
  \bibinfo{pages}{573} (\bibinfo{year}{1975}).

\bibitem[{\citenamefont{Burgarth and Giovannetti}(2007)}]{Burgarth2007c}
\bibinfo{author}{\bibfnamefont{D.}~\bibnamefont{Burgarth}} \bibnamefont{and}
  \bibinfo{author}{\bibfnamefont{V.}~\bibnamefont{Giovannetti}}
  (\bibinfo{year}{2007}), \bibinfo{note}{proceedings, M. Ericsson and S.
  Montangero (eds.), Pisa, Edizioni della Normale 2008 (arXiv:0710.0302)}.

\bibitem[{\citenamefont{Agranovitch}(2008)}]{Agranovich2008}
\bibinfo{author}{\bibfnamefont{V.~M.} \bibnamefont{Agranovitch}},
  \emph{\bibinfo{title}{Excitations in Organic Solids}}
  (\bibinfo{publisher}{Oxford University Press}, \bibinfo{year}{2008}).

\bibitem[{\citenamefont{Ishizaki and Fleming}(2009)}]{Ishizaki2009}
\bibinfo{author}{\bibfnamefont{A.}~\bibnamefont{Ishizaki}} \bibnamefont{and}
  \bibinfo{author}{\bibfnamefont{G.~R.} \bibnamefont{Fleming}},
  \bibinfo{journal}{Proc. Natl. Acad. Sci. USA} \textbf{\bibinfo{volume}{106}},
  \bibinfo{pages}{17255} (\bibinfo{year}{2009}).

\bibitem[{\citenamefont{Hoyer et~al.}(2010)\citenamefont{Hoyer, Sarovar, and
  Whaley}}]{Hoyer-etal:NJP2010}
\bibinfo{author}{\bibfnamefont{S.}~\bibnamefont{Hoyer}},
  \bibinfo{author}{\bibfnamefont{M.}~\bibnamefont{Sarovar}}, \bibnamefont{and}
  \bibinfo{author}{\bibfnamefont{K.~B.} \bibnamefont{Whaley}},
  \bibinfo{journal}{New J. Phys.} \textbf{\bibinfo{volume}{12}},
  \bibinfo{pages}{065041} (\bibinfo{year}{2010}).

\bibitem[{\citenamefont{Schirmer et~al.}(2008)\citenamefont{Schirmer, Oi, and
  Devitt}}]{Schirmer2008}
\bibinfo{author}{\bibfnamefont{S.}~\bibnamefont{Schirmer}},
  \bibinfo{author}{\bibfnamefont{D.}~\bibnamefont{Oi}}, \bibnamefont{and}
  \bibinfo{author}{\bibfnamefont{S.}~\bibnamefont{Devitt}},
  \bibinfo{journal}{Inst. Phys. Conf. Ser.} \textbf{\bibinfo{volume}{107}},
  \bibinfo{pages}{012011} (\bibinfo{year}{2008}).

\bibitem[{\citenamefont{Brivio et~al.}(2010)\citenamefont{Brivio, Cialdi,
  Vezzoli, Gebrehiwot, Genoni, Olivares, and Paris}}]{Brivio2010}
\bibinfo{author}{\bibfnamefont{D.}~\bibnamefont{Brivio}},
  \bibinfo{author}{\bibfnamefont{S.}~\bibnamefont{Cialdi}},
  \bibinfo{author}{\bibfnamefont{S.}~\bibnamefont{Vezzoli}},
  \bibinfo{author}{\bibfnamefont{B.~T.} \bibnamefont{Gebrehiwot}},
  \bibinfo{author}{\bibfnamefont{M.~G.} \bibnamefont{Genoni}},
  \bibinfo{author}{\bibfnamefont{S.}~\bibnamefont{Olivares}}, \bibnamefont{and}
  \bibinfo{author}{\bibfnamefont{M.~G.~A.} \bibnamefont{Paris}},
  \bibinfo{journal}{Phys. Rev. A} \textbf{\bibinfo{volume}{81}},
  \bibinfo{pages}{012305} (\bibinfo{year}{2010}).

\bibitem[{\citenamefont{Schirmer and Oi}(2010)}]{Schirmer2010}
\bibinfo{author}{\bibfnamefont{S.}~\bibnamefont{Schirmer}} \bibnamefont{and}
  \bibinfo{author}{\bibfnamefont{D.}~\bibnamefont{Oi}}, \bibinfo{journal}{Laser
  Phys.} \textbf{\bibinfo{volume}{20}}, \bibinfo{pages}{1203}
  (\bibinfo{year}{2010}).

\end{thebibliography}

\end{document}